\documentclass{aastex}
\usepackage{emulateapj5}

\begin{document}

\title{The Stellar Initial Mass Function in Primordial Galaxies}

\author{Fumitaka Nakamura}
\affil{Faculty of Education and Human Sciences, Niigata University,
8050 Ikarashi-2, Niigata 950-2181, Japan, and Astronomy Department, 
University of California, Berkeley, Berkeley, CA 94720}
\and
\author{Masayuki Umemura}
\affil{Center for Computational Physics, University of Tsukuba,
Tsukuba, Ibaraki 305-8577, Japan}

\begin{abstract}

In the context of star formation through fragmentation of
an extremely metal-deficient protogalactic cloud,
the gravitational collapse of filamentary gas clouds is explored with
one-dimensional numerical hydrodynamics coupled with 
non-equilibrium chemistry of H$_2$ and HD.
It is found that the cloud evolution is governed mainly by
the initial central density ($n_{\rm c,0}$) 
and H$_2$ abundance ($x_{\rm H_2,0}$).
In particular, the evolution of low-density filaments 
($n_{\rm c,0} \lesssim 10^5$ cm$^{-3}$) 
bifurcates at a threshold H$_2$ abundance
 of $x_{\rm H_2,cr}\simeq 3\times 10^{-3}$,
 beyond which HD cooling overwhelms H$_2$ cooling.
The contraction of a filament with $n_{\rm c,0} \lesssim 10^5$ cm$^{-3}$ and
 $x_{\rm H_2, 0}\gtrsim x_{\rm H_2,cr}$ is strongly decelerated 
when the central density ($n_{\rm c}$) reaches a critical density of HD
at which LTE level populations are achieved,
and therefore the filament is
expected to fragment at $\sim 10^{7}$ cm$^{-3}$.
The fragment mass is assessed to be $\approx 10M_\odot$.
In contrast, the contraction of a filament with 
$n_{\rm c,0} \lesssim 10^5$ cm$^{-3}$ and
 $x_{\rm H_2, 0}\lesssim x_{\rm H_2,cr}$
is regulated by H$_2$ cooling.
In this case, the filament tends to fragment at lower density
as $\sim 10^{4}$ cm$^{-3}$ owing to the low critical density of H$_2$, 
and the fragment mass is as high as $\approx 10^2M_\odot$.
For a high-density filament with $n_{\rm c,0} \gtrsim 10^5$ cm$^{-3}$,
the temperature stays at a relatively high value 
 because both H$_2$ and HD cooling saturate,
and the cloud evolution is governed by H$_2$ cooling.
The contraction of a high-density filament is accelerated
 by effective three-body H$_2$ formation when the density reaches
 $10^{8-9}$ cm$^{-3}$.
The fragmentation is not expected to take place until the cloud
becomes opaque in H$_2$ lines at $n_{\rm c,0}\sim 10^{12-13}$ cm$^{-3}$, 
so that the fragment mass is reduced to $1-2$ M$_\odot$.
As a result, the stellar initial mass function (IMF) could be 
bimodal and deficient in sub-solar mass stars,
where the high mass peak is around $10M_\odot$ or $10^2M_\odot$,
dependently on $n_{\rm c,0}$ and $x_{\rm H_2,0}$.
If the protogalactic clouds are ionized by UV radiation or
strong shocks, the H$_2$ abundance could exceed
 $x_{\rm H_2,cr}\simeq 3\times 10^{-3}$
by reactions of $\rm H+e \rightarrow H^- + {\it h\nu}$ and
$\rm H+H^-\rightarrow H_2+e$.
The high mass peak would then be $O(10) M_\odot$.

\end{abstract}


\section{Introduction}

The first generation of stars, Population III stars, are
thought to form from almost metal-free gas with metallicity of 
 $\sim 10^{-10} Z_\odot$ (see e.g., Carr, Bond, \&
Arnett 1984; Carr 1994).
Such Population III stars may have produced heavy elements
at redshifts $z \gtrsim 10$.
Recent observations of quasar absorption spectra show that
the metal abundance in intergalactic space is 
$\approx 10^{-3} Z_\odot$ (e.g., Cowie \& Songaila 1998). 
Also, it is widely accepted that a significant portion of old halo 
stars in our Galaxy have metallicity lower than $10^{-3}Z_\odot$
(Beers, Preston, \& Shectman 1992).
In addition, some blue compact dwarf galaxies are known
to be extremely metal-poor ($\approx 10^{-2} Z_\odot$, 
Kunth \& Sargent 1986; Pustilnik et al. 2001).
Thus, star formation in protogalaxies must have proceeded
in very metal-deficient environments.

When the metallicity is lower than $Z \lesssim 10^{-2}Z_\odot$, 
the cooling by heavy elements is less effective than cooling 
by hydrogen and helium (e.g., Yoshii \& Sabano 1980; 
B\"ohringer \& Hensler 1989; Omukai 2000; Nishi \& Tashiro 2000).
For the formation of Population III stars from such metal-deficient gas, 
the cooling by primordial hydrogen molecules (H$_2$) 
plays an essential role
 (Matsuda, Sato, \& Takeda 1969; Yoneyama 1972;
 Hutchins 1976; Silk 1977; Yoshii \& Sabano 1980; 
 Carlberg 1981; Lepp \& Shull 1984; Palla, Salpeter, \& Stahler 1983;
 Yoshii \& Saio 1986; Shapiro \& Kang 1987;
 Uehara et al. 1996; 
 Haiman, Thoul \& Loeb 1996; Nishi et al. 1998; 
 Abel et al. 1998; Bromm, Coppi, \& Larson 1999; 
 Abel, Bryan, \& Norman 2000; Nakamura \& Umemura 1999a, 2001, hereafter 
Papers I and II).
Recent studies have revealed that star formation in primordial gas
is considerably different from present-day star formation, implying that 
the stellar initial mass function (IMF) might be different 
from the Salpeter-like IMF and 
the IMF might be time-varying in the course of 
galaxy evolution from the early collapsing stages to 
the present day (see also Larson 1998; Zepf \& Silk 1996; Chabrier 1999).

Besides hydrogen molecules,
deuterated hydrogen molecules (HD) can be a significant coolant
(e.g., Galli \& Palla 1998).
Although HD is less abundant than H$_2$
([HD/H$_2$] $\sim 10^{-3}- 10^{-4}$), HD has
a finite dipole moment and thus higher
radiative transition probabilities than H$_2$.
For example, the lowest rotational transitions of H$_2$ and HD
have radiative transition probabilities of 
$A_{20}=3\times 10^{-11}$ s$^{-1}$ and 
$A_{10}=5\times 10^{-8}$ s$^{-1}$, respectively. 
Also, the corresponding excitation energies of H$_2$ and HD are
$\Delta E_{20}/k = 510$ K and $\Delta E_{10}/k = 128$ K, respectively, 
where $k$ is the Boltzmann coefficient.
Thus, HD can lower the gas temperature down to $T\lesssim$ 100K 
(e.g., Puy \& Signore 1996; Bougleux \& Galli 1997;
 Galli \& Palla 1998; Flower et al. 2000).

The first pregalactic objects should have collapsed at redshifts
 of $z\sim 10 - 10^2$ and have masses of $10^5 - 10^8$ M$_\odot$ 
in a cold dark matter cosmology
 (Tegmark et al. 1998; Fuller \& Couchman 2000). 
When such first objects are virialized and
the gas temperature ascends to $10^3 - 10^4$ K, 
H$_2$ formation is promoted and the fractional abundance is raised to
$x_{\rm H_2}=10^{-4} - 10^{-3}$.
However, for this H$_2$ abundance,
the H$_2$ cooling cannot lower the temperature down to 100 K.
Thus, the cloud evolution is basically controlled by H$_2$ cooling
and HD cooling is not important
(Nakamura \& Umemura 1999b, 2000; see also Lepp \& Shull 1984).

But, in some situations, the H$_2$ abundance can be raised further.
Ferrara (1998) argued that in dense shells formed behind supernova shocks, 
 H$_2$ molecules form efficiently owing to reduced recombination 
 of free electrons (see also Shapiro \& Kang 1987).  
In the dense shells, the H$_2$ concentration reaches $6 \times 10^{-3}$, 
and consequently the gas temperature descends to $\sim 100$ K.
At such low temperature, HD is a more efficient coolant than H$_2$.
The primordial star formation regulated by HD cooling 
in shocked shells is studied by several authors
 (Uehara \& Inutsuka 2000; Machida, Fujimoto, \& Nakamura 2001).
Another possibility is star formation from photoionized gas.
Corbelli, Galli, \& Palla (1998) studied the effects of
UV background radiation on the thermal evolution of
 the protogalaxies
 (see also Murray \& Lin 1989; Haiman, Rees, \& Loeb 1996).  
They found that at redshifts lower than $z \sim 1-2$,
protogalactic clouds are self-shielded against 
the UV background radiation, and 
H$_2$ formation is promoted up to $x_{\rm H_2} \approx 10^{-2}$
with the help of abundant free electrons due to the UV radiation.
Even at higher redshifts ($z\gtrsim 2$), if pregalactic clouds more
 massive than $10^{11} M_\odot$ collapse under a UV background,
they can be eventually self-shielded and abundant H$_2$ ($\gtrsim 10^{-3}$)
forms, so that the temperature descends down to $\sim 100$K (Susa \& Umemura 2000).
Hence, HD cooling is likely to be important for the star formation
 in the protogalactic systems.
In this paper, we extensively examine the effects of the HD cooling
on the formation of stars in protogalaxies, and 
elucidate the role of HD molecules for the stellar initial mass function
(IMF) there.

This paper is organized as follows.  In \S 2, we describe 
model and numerical methods which are basically the same as
those of Paper II.  Numerical results are presented in \S 3.
In \S 4, implications for the IMF of Population III and
metal-deficient stars are discussed.

\section{Model and Numerical Methods}
\label{sec:model}

In Papers I and II, we studied the fragmentation
of filamentary clouds by incorporating the H$_2$ cooling.
In this paper, we examine the effects of HD cooling
on the evolution of the filaments.
Our numerical model and method are the same as
those of Paper II except for the inclusion
of deuterium chemistry.
In the following, we briefly review our numerical model.
See \S 2 of Paper II for more detail.

We deal with the following 14 species:
e, H, H$^+$, H$^-$, H$_2$, H$_2^+$, He, He$^+$, He$^{++}$,
D, D$^+$, D$^-$, HD, and HD$^+$.
The mass fraction of He is set to 0.24 of the total mass. 
The D abundance is set to $4\times 10^{-5}$ by number,  
which is consistent with recent observations
of the deuterium Ly-$\alpha$ feature in the absorption spectra 
of high-redshift quasars (e.g., Tytler et al. 1996; O'Meara et al. 2000).
The reaction rate coefficients for the deuterium chemistry are 
given in Table \ref{tab:hd rate},
while the rate coefficients for other species
are the same as those of Paper II.

We take into account the following thermal processes: 
(1) H cooling by radiative recombination, 
collisional ionization, and collisional excitation, 
(2) H$_2$ line cooling by rotational and
vibrational transitions, 
(3) cooling by H$_2$ collisional dissociation, 
(4) heating by H$_2$ formation, and 
(5) HD line cooling by rotational transitions.

The HD line cooling is computed by the same method 
as the H$_2$ line cooling described in Paper II, which
includes the escape probability method in an optically-thick
regime.  For the collisional deexcitation rates for HD,
we consider both H-HD and H$_2$-HD collisions, 
using analytical fits of Galli \& Palla (1998) and 
Flower \& Roueff (1999).
The first 6 rotational levels are taken into account.
Recently, Flower et al. (2000) derived an updated HD cooling function
in an optical thin regime, which includes up to 8 rotational levels.
We have checked the difference by employing their cooling function 
for several models described in \S 3,
and found that the resultant temperatures agree with 
the present results to within $\sim$4\%.

We consider an infinitely long cylindrical gas cloud
which collapses in the radial direction. 
The initial temperature is assumed to be spatially constant.
The initial abundances of electron and D$^+$ are set to 
$x_e=5\times 10^{-5}$ and $x_{\rm D^+}=1\times 10^{-7}$,
respectively.
(The numerical results do not depend sensitively 
upon $x_e$ and $x_{\rm D^+}$. See also \S 2 of Papers I and II.)
At the initial state, the relative abundances of 
H$^-$, H$_2^+$, He$^+$, He$^{++}$, D$^{-}$, HD, and HD$^{+}$
are set to zero for simplicity.
The abundances of other species are determined by the conservation of 
mass and charge.
The initial density profile in the radial direction is assumed to be 
\begin{equation}
\rho_0 =\rho _{\rm c,0} \left(1+r^2/R_{0}^2\right)^{-2} \, ,
\end{equation}
where $R_0=\sqrt{2fkT_0/(\pi G\rho _{\rm c,0}\mu)}$ is the effective radius,
$\rho_{\rm c,0}$ is the central mass density, $T_0$ is the initial
gas temperature, $\mu$ is the mean molecular weight, 
and $f$ is the ratio of the gravitational force
to the pressure force.
When $f=1$, the density distribution agrees with
that of an isothermal cylinder in hydrostatic equilibrium.
When the filamentary cloud forms through gravitational fragmentation 
of a parent sheet-like cloud, $f$ is expected to be $\approx 2$
(see \S 2.2 of Paper I).
The radial infall velocity is given by
\begin{equation}
v_{r, 0} =-v_0 r\left(R_0 +\sqrt{R_0^2+r^2}\right)^{-1}\, ,
\end{equation}
where $v_0$ is constant and is set to the initial sound speed. 
We neglect the effects of dark matter because after virialization
of the parent system, the local density of baryonic gas
is likely to be higher than the background 
dark matter density owing to radiative cooling
(e.g., Umemura 1993).
The model is thus specified by four parameters: $n_{\rm c,0}$,
$T_0$, $f$, and the initial H$_2$ fractional abundance $x_{\rm H_2, 0}$.
As stated in \S1, high H$_2$ abundance as 
$x_{\rm H_2, 0} \gtrsim 10^{-3}$ 
can be achieved if
a parent cloud undergoes ionization prior to the collapse.

In Paper II, we followed the collapse and fragmentation of the
primordial filaments with one-dimensional and two-dimensional 
axisymmetric simulations.
The numerical results of Paper II showed that 
the fragment masses derived from the two-dimensional simulations 
are in good agreement with the estimate based on one-dimensional
simulations. 
Therefore, in this paper, we pursue one-dimensional simulations 
on the collapse of the filaments to estimate the fragment masses.

We calculated more than 3000 models by choosing the model parameters
$n_{\rm c,0}$, $T_0$, $f$, and $x_{\rm H_2,0}$ 
as $\log _{10} (n_{\rm c,0}/{\rm cm}^{-3}) 
= 1.00, 1.33, 1.67, 2.00, \dots, 6.00$, 
$T_0=200$, 300, 400 K, $f=1.5, 2.0, 2.5, \dots, 6.0$, and
$x_{\rm H_2,0}= 1\times 10^{-4}, 3\times 10^{-4}, 6\times 10^{-4},
1\times 10^{-3}, \dots, 1\times 10^{-2}$, respectively.

\section{Numerical Results}

In Papers I and II, we studied the fragmentation of the 
primordial filaments incorporating H$_2$ cooling,
and it was shown that the fragmentation is bifurcated by 
a threshold initial density of $\approx 10^5{\rm cm}^{-3}$, 
below which as massive stars as $\approx 10^2M_\odot$ form, and
above which less massive stars with $\approx 1M_\odot$ form.

In this section, we reexamine the collapse of the filaments, 
including the HD cooling as well as H$_2$ cooling.
As shown below, there is a threshold initial H$_2$ concentration,
above which HD cooling predominantly regulates the cloud evolution. 
We also show that the HD cooling does not play an important role
 for high-density gas.
Thus, the evolution of the filaments is classified
into three cases, depending upon the initial density, $n_{\rm c,0}$,
and initial H$_2$ abundance, $x_{\rm H_2, 0}$; 
(1) low-density filaments with high $x_{\rm H_2, 0}$, (2) low-density
filaments with low $x_{\rm H_2, 0}$, and (3) high-density filaments.

In the following, we first compare the HD cooling and H$_2$ cooling
 on the $n_{\rm c,0}-x_{\rm H_2, 0}$ diagram to clarify
 the importance of HD cooling.
Next, we show the numerical results of one-dimensional simulations.
In the models shown in this section, the initial temperatures
 are set to be equilibrium temperatures of the filaments which
decrease with increasing $x_{\rm H_2, 0}$.

\subsection{Threshold H$_2$ Abundance}

\begin{figure}
\epsscale{0.5}
\plotone{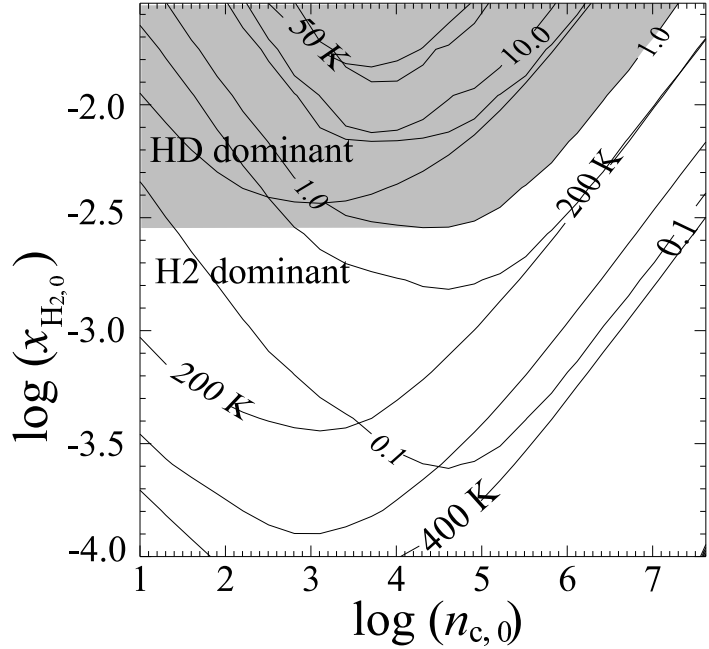}
\caption{Comparison between the HD cooling rate and H$_2$ cooling rate.
The figure shows the HD-to-H$_2$ cooling ratio 
in the initial gas density ($n_{\rm c,0}$) $-$ 
initial H$_2$ abundance ($x_{\rm H_2, 0}$) diagram.
The HD abundance is assumed to be $1\times 10^{-4}$ times the H$_2$ abundance.
The solid lines show the contour curves of the HD-to-H$_2$
 cooling ratio, where the equilibrium gas temperatures are
 determined by the condition of $t_{\rm frag}=t_{\rm cool}$
 and are indicated by dashed lines.
If the initial parameters of the filaments 
 are in the gray region, the HD cooling plays an important
 role in the thermal evolution of the gas (see the text for more details).
\label{fig:1}}
\end{figure}

HD cooling depends sensitively on the initial 
cloud density and H$_2$ abundance.
To clarify the effect of HD cooling, 
we compare the HD cooling rate ($\Lambda _{\rm HD}$) and 
 the H$_2$ cooling rate ($\Lambda _{\rm H_2}$) in Figure \ref{fig:1}. 
The abscissa and ordinate indicate the initial
gas density and the H$_2$ abundance, respectively.
The solid lines show the contour curves of
the HD-to-H$_2$ cooling ratio ($\Lambda _{\rm HD}$/$\Lambda _{\rm
H_2}$). 
When evaluating the cooling rates, the gas temperatures 
are determined iteratively to satisfy 
the condition of $t_{\rm cool}=t_{\rm frag}$,
 where $t_{\rm cool}$ is cooling time and $t_{\rm frag}$ is 
 fragmentation time as defined in \S \ref{subsec:LowD1}.
The resultant temperatures are also shown by dashed lines.
The HD abundance is taken to be proportional
 to the H$_2$ abundance as $x_{\rm HD}/x_{\rm H_2}=1\times 10^{-4}$,
which is consistent with those in the models shown
 in \S \ref{subsec:LowD2} and \ref{subsec:HighD}.
This figure shows that
the contribution of HD cooling rate is largest
 around $n_{\rm c,0}=10^5$ cm$^{-3}$,
 which is almost comparable to the critical density of HD,
 beyond which the rotational level populations achieve the LTE. 

If the cloud has the initial H$_2$ abundance
 lower than a threshold value of $x_{\rm H_2, cr}=3\times 10^{-3}$, 
 HD cooling does not play a significant role
 in the thermal evolution of the cloud
 because $\Lambda _{\rm HD}<\Lambda _{\rm H_2}$.
The cloud temperature is then equal to the equilibrium temperature 
shown by the dashed lines.
As shown later, the H$_2$ abundance in a collapsing filament stays 
almost constant before the three-body reactions proceed at $10^{8-9}$ cm$^{-3}$. 
Thus, if an evolutionary path of such a cloud is 
superimposed on Figure 1, then it is almost parallel to the abscissa.
For the filaments with 
 $n \lesssim 10^5$ cm$^{-3}$ and $x_{\rm H_2} \gtrsim x_{\rm H_2, cr}$, 
the evolutionary path enters into the region of
 $\Lambda _{\rm HD}>\Lambda _{\rm H_2}$ in the course of contraction even if 
 $\Lambda _{\rm H_2}$ is larger than $\Lambda _{\rm HD}$
 at the initial state.
In actual evolution, as shown below, the pressure force 
retards the contraction of the filament.
In particular, when the density exceeds HD critical density, 
$n_{\rm HD, cr}\sim 10^{4-5}$ cm$^{-3}$, 
the contraction slows 
and the actual cloud temperature stays below 100 K.
Consequently, HD cooling continues to control 
the contraction until the cloud becomes opaque to the HD lines.
On the other hand, for the filaments with 
$n_{\rm c,0} \gtrsim 10^5$ cm$^{-3}$ 
and $x_{\rm H_2,0} \gtrsim x_{\rm H_2, cr}$,
 the evolutionary path goes into the region of 
 $\Lambda _{\rm HD}<\Lambda _{\rm H_2}$  even if 
 $\Lambda _{\rm HD}$ is larger than $\Lambda _{\rm H_2}$ at 
 the initial state.
As a result, for the initial condition in the gray region in Figure 1, 
the HD cooling plays an important
 role in the thermal evolution of the gas.

\subsection{Low-Density Filaments with high H$_2$ abundance
($n_{\rm c,0}\lesssim 10^5 {\rm cm^{-3}}$ and 
$x_{\rm H_2, 0}\gtrsim 3\times 10^{-3}$)}
\label{subsec:LowD1}

As a typical example in which the HD cooling controls the contraction, 
we show the evolution of the model
with $n_{\rm c,0}=10$ cm$^{-3}$, $T_0=200$ K, $f=2$, and 
$x_{\rm H_2, 0} = 3 \times 10^{-3}$.

\begin{figure}
\epsscale{0.5}
\plotone{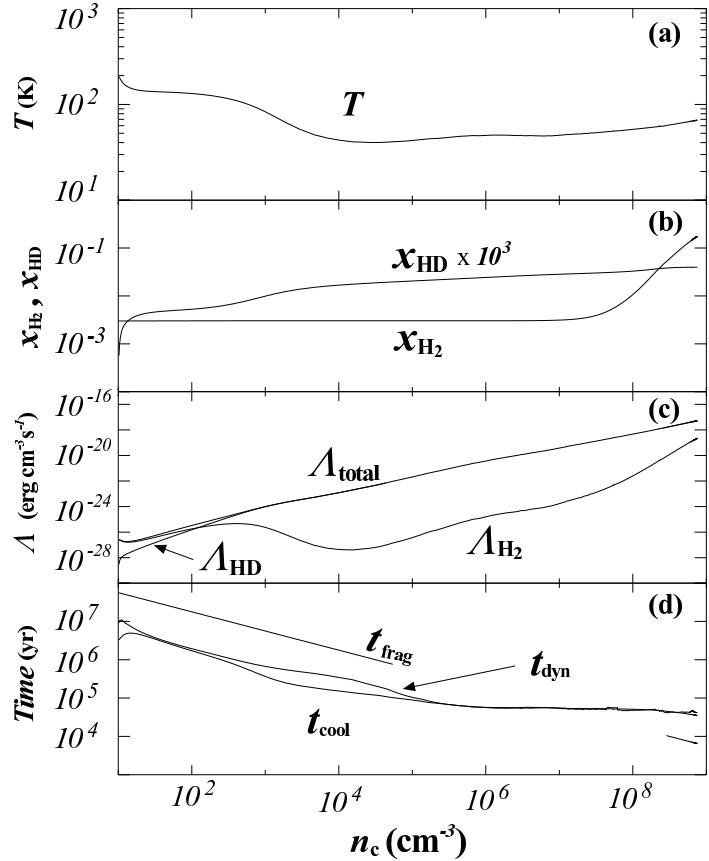}
\caption{Evolution of (a) the temperature,
 (b) H$_2$ ({\it solid line}) and HD ({\it dashed line}) abundances, 
(c) total ({\it solid line}), H$_2$ ({\it dashed line}),
 and HD ({\it dotted line}) cooling rates, (d) the cooling time 
 ({\it solid line}), the contraction time  ({\it dashed line}),
 and the fragmentation time ({\it dotted line}), respectively, 
 as a function of the central density [$n_c\equiv \rho_c /(\mu m_H) $]
 for the model with $(n_{\rm c,0}, T_0, f, x_{\rm H_2, 0})=
 (10 \, {\rm cm}^{-3},\ 200 \, {\rm K}, \ 2, \ 3 \times 10^{-3})$. 
Note that the definition of the fragmentation time is 2.5 times larger
 than that in Paper II.
\label{fig:2}}
\end{figure}

Figures \ref{fig:2}a, \ref{fig:2}b, \ref{fig:2}c, and \ref{fig:2}d show
the time evolution of (a) the temperature,
 (b) H$_2$ ({\it solid line}) and HD ({\it dashed line}) abundances, 
(c) cooling rates by H$_2$ ({\it dashed line})
 and by HD ({\it dotted line}), and the total ({\it solid line}),
(d) the cooling timescale
 ({\it solid line}), the timescale of dynamical contraction
 ({\it dashed line}),
 and the fragmentation timescale ({\it dotted line}), respectively, 
 as a function of the central density [$n_c\equiv \rho_c /(\mu m_H) $].
Since the central density monotonically increases with time,
the abscissa corresponds to the evolution time.
The contraction and cooling timescale are, respectively, defined as 
$t_{\rm dyn}\equiv \rho/\dot{\rho}$ and 
$t_{\rm cool}\equiv E_{\rm t}/(\Lambda _{\rm t})$, 
where $E_{\rm t}$ and $\Lambda _{\rm t}$ are total 
internal energy and net cooling rate, respectively. 
The fragmentation time, $t_{\rm frag}$, is defined as 2.5 times the inverse of 
the growth rate of the fastest-growing mode
(eq [38] of Nakamura, Hanawa, \& Nakano 1993), i.e., 
$t_{\rm frag} \equiv 5.17 (2 \pi G \rho)^{-1/2}$.
Note that this definition of $t_{\rm frag}$ 
 is 2.5 times larger than that of Paper II.
In practice, a filament does not fragment in
one-dimensional calculations. Therefore, $t_{\rm frag}$ should be
regarded as a measure of the fragmentation epoch as well as 
the free-fall time-scale.

At the early stages of the contraction, H$_2$ cooling dominates
HD cooling, and the equilibrium temperature 
is lowered to $T\sim 100$ K owing to the high H$_2$ abundance.
When the density reaches the critical density of H$_2$,
$n_{\rm H_2, cr}\sim 10^{3-4}$ cm$^{-3}$,
the H$_2$ cooling is saturated and then HD cooling becomes dominant as the
density increases.
When the HD cooling becomes effective, the equilibrium temperature
descends to $\sim $50 K owing to the lower excitation temperatures of HD.
Thereafter, the temperature stays nearly constant at $\sim$50 K
during the contraction.
When the density reaches 
$n_{\rm HD,cr} \sim 10^{4-5}$ cm$^{-3}$,
the cloud contraction tends to become quasistatic
by the combination effect of
(1) the dynamical stability of a {\it cylindrical} cloud and (2)
the characteristic of line cooling.
First, a polytropic cylinder in hydrostatic equilibrium 
with $P\propto \rho ^\gamma$ is gravitationally stable in the
radial directions,
if the effective $\gamma$ is greater than unity.
This means that an isothermal cylinder is marginally stable.
Second, at lower density than the critical density,
the line cooling rate is nearly proportional to $n^2$
because the collisional excitation rate balances
with spontaneous emission rate,
while at higher density than the critical density,
the cooling rate is proportional to $n$ because the level populations
reach LTE. 
Hence, $t_{\rm cool}$ is inversely proportional to the density
for $n<n_{\rm HD,cr}$, while $t_{\rm cool}$ is constant for $n>n_{\rm HD,cr}$
if the temperature is constant.
By these effects, the compressional heating strongly brakes the 
dynamic contraction when $n>n_{\rm HD,cr}$ and 
then the cloud evolves quasistatically.

When the density reaches $\sim 10^9$ cm$^{-3}$,
the cloud becomes optically thick in the HD lines
and the quasistatic contraction almost ceases.
However, before the central density reaches this density, 
the filament is likely to fragment.
As shown by two-dimensional calculations in Paper II,
it is anticipated that 
the fragmentation takes place when the dynamical timescale
becomes a few times longer than the fragmentation timescale
(see Paper II for more detail). 
From Figure 2d, this cloud is expected to undergo the fragmentation when the
density reaches 10$^7$ cm$^{-3}$.

Uehara \& Inutsuka (2000) estimated the minimum fragment mass
when the HD lines become opaque. 
However, in their model, the cloud becomes optically 
thick to the HD lines when the density reaches 
$\approx 3\times 10^{10}$ cm$^{-3}$,
which is about 30 times higher than the present result.
To understand this discrepancy, we compare our evaluation of 
the optical depth with the LVG approximation 
(Goldreich \& Kwan 1974) with some modification.
The optical depth of the LVG approximation is originally given by 
\begin{equation}
 \tau _{J+1,J} = \frac{hc}{4\pi}\frac{B_{J,J+1}n_J}{|dv/ds|}\left(
  1-\frac{g_J n_{J+1}}{g_{J+1}n_J} \right) \; ,
\end{equation}
where $h$ is the Plank constant, $c$ is the speed of light,
 $B_{J,J+1}$ is Einstein's B coefficient of the transition
 $J\rightarrow J+1$, $|dv/ds|$ is the velocity gradient, 
 $n_J$ and $g_J$ are the number density and 
 the Gaunt factor of the $J$ level, respectively.
However, in the present calculations,
the velocity gradient is small compared to the thermal width
and the gradient of the thermal velocity is dominant.
Hence, we replace the velocity gradient by
 $\alpha v_{\rm th}/R_{\rm J} = \alpha \sqrt{\pi G\rho _c}$,
 where $\alpha$, $v_{\rm th}$, and $R_{\rm J}$ are a
 nondimensional numerical constant, the thermal velocity,
 and the filament radius, respectively.
If we set the temperature to 60 K, the optical depth of the 
 most effective cooling line ($J=1\rightarrow 0$)
 becomes unity when the density reaches 
$1- 3 \times 10^{9}$ cm$^{-3}$ for $\alpha = 1-2$.
This is consistent with the present result.
Also, it should be noted that our definition of the fragment mass is
 different from that of Uehara \& Inutsuka (2000).
We adopt the mass contained within
 one wavelength of the fastest-growing linear perturbation
 which is consistent with our two-dimensional numerical results
 (see \S \ref{sec:dependence}), while
 Uehara \& Inutsuka (2000) adopted the Jeans mass which is about ten
 times smaller than our definition at the same density and temperature.
As a result, their estimation of the fragment mass
is by about two orders of magnitude smaller than our estimate
if we assume the cloud to fragment at the stage 
optically thick to the HD lines.
However, in practice, the cloud is likely to fragment at the earlier stages
around the critical density of HD as shown above.

The evolution of the other models with different initial parameters 
($10$ cm$^{-3}$ $\lesssim n_{\rm c,0} \lesssim $ $10^4$ cm$^{-3}$,
 200 K $\lesssim T_0 \lesssim $ 400 K, $1.5\lesssim f \lesssim 6$, and
 $3\times 10^{-3} \lesssim x_{\rm H_2, 0}\lesssim 10^{-2}$)
is qualitatively similar to that of the model shown in Figure \ref{fig:2}.
When the initial H$_2$ abundance is as high as $10^{-2}$, 
the HD cooling is more effective than H$_2$ cooling at the initial state.
Thus, the temperature goes down to $\sim 50$ K at the very early
stages of the evolution.

\subsection{Low-Density Filaments with Low H$_2$ Abundances
($x_{\rm H_2, 0}\lesssim 3\times 10^{-3}$)}
\label{subsec:LowD2}

\begin{figure}
\epsscale{0.5}
\plotone{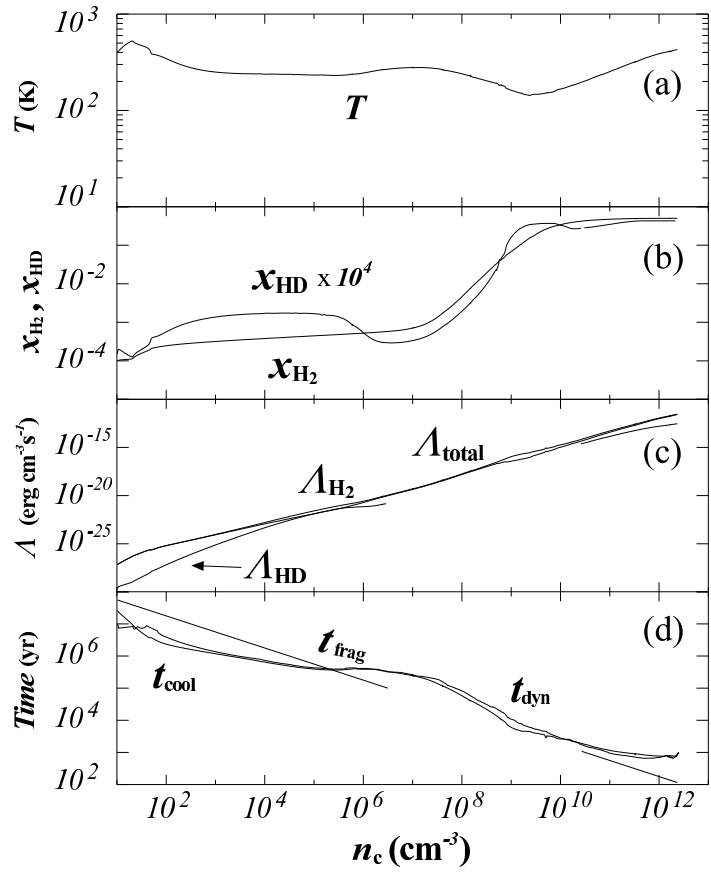}
\caption{Same as Figure \ref{fig:2} but for the model
 with $(n_{\rm c,0}, T_0, f, x_{\rm H_2, 0})= (10 \, {\rm cm}^{-3},
 \ 400 \, {\rm K}, \ 2, \ 1 \times 10^{-4})$.
\label{fig:3}}
\end{figure}

As a typical example of this case, we show the evolution of the model
with $n_{\rm c,0}=10$ cm$^{-3}$, $T_0=400$ K, $f=2$, and 
$x_{\rm H_2, 0} = 1 \times 10^{-4}$ in Figure \ref{fig:3}.
The notation of the figure is the same as that of Figure \ref{fig:2}. 

In this model, $\Lambda _{\rm HD} < \Lambda _{\rm H_2}$
during the contraction, and thus
HD cooling does not play a significant role
in the thermal and dynamical evolution.
The temperature remains between 100 and 500 K over 11
orders of magnitude in density until the H$_2$ lines become optically thick.
During the contraction, the HD abundance is nearly proportional
to that of H$_2$ ($x_{\rm HD}\sim 10^{-4}x_{\rm H_2}$)
because HD molecules form primarily via the reaction 
${\rm D + H_2 \rightarrow H + HD}$.
Thus, when the three-body H$_2$ formation becomes effective,
almost all the D atoms are processed into HD.
It should be noted that the above reaction is sensitive to
the temperature, and for lower temperature, the reactions 
${\rm D^+ + H_2 \rightarrow H^+ + HD}$ and 
${\rm HD^+ + H \rightarrow D^+ + HD}$ are more effective.

The contraction proceeds quasistatically
after the density reaches the critical density of H$_2$
($n_{\rm H_2, cr}=10^{3-4}$ cm$^{-3}$).
When the three-body H$_2$ formation becomes effective, 
the contraction is accelerated again 
and then decelerated when the cloud becomes
optically thick to the H$_2$ lines ($n_{\rm c} \sim 10^{12}$ cm$^{-3}$).
The cloud is expected to fragment after the density reaches the
critical density of H$_2$.

The evolution of the other models with different parameters 
($10$ cm$^{-3}$ $\lesssim n_{\rm c,0} \lesssim $ $10^4$ cm$^{-3}$,
 200 K $\lesssim T \lesssim $ 400 K, $1.5\lesssim f \lesssim 6$, and
 $10^{-4} \lesssim x_{\rm H_2, 0}\lesssim 10^{-3}$)
is qualitatively similar to that of this model.

\subsection{High-Density Filaments}
\label{subsec:HighD}

\begin{figure}
\epsscale{0.5}
\plotone{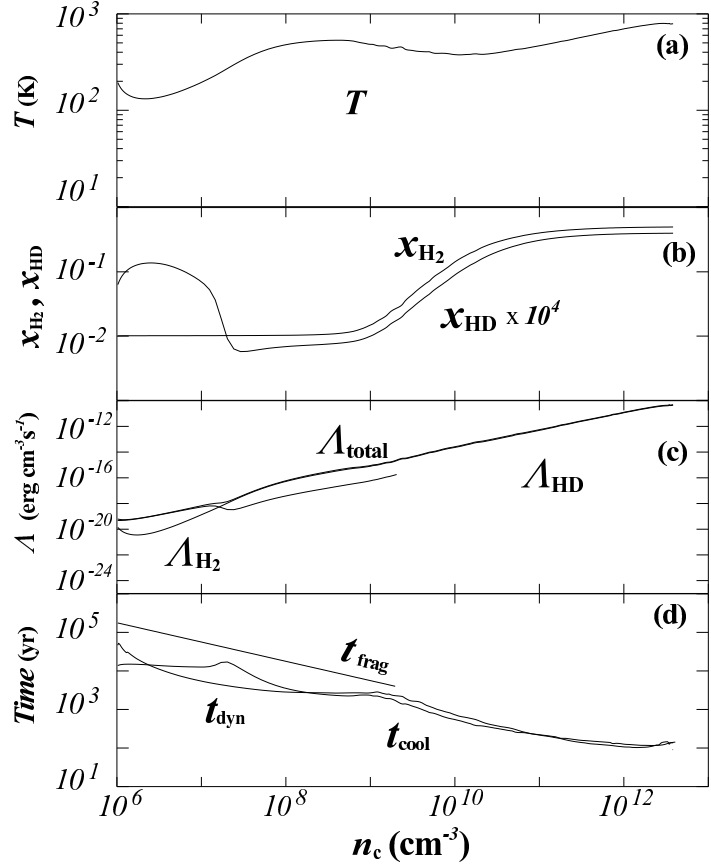}
\caption{Same as Figure \ref{fig:2} but for the model
 with $(n_{\rm c,0}, T_0, f, x_{\rm H_2, 0})= (10^6 \, {\rm cm}^{-3},
 \ 200 \, {\rm K}, \ 6, \ 1 \times 10^{-2})$.
\label{fig:4}}
\end{figure}

As a typical example of this case, we show the evolution of the model
with $n_{\rm c,0}=10^6$ cm$^{-3}$, $T_0=200$ K, $f=6$,
and $x_{\rm H_2, 0}=1\times 10^{-2}$ in Figure \ref{fig:4}. 
The notation of the figure is the same as that of Figure \ref{fig:2}.

Although the initial H$_2$ abundance is as high as the threshold H$_2$
abundance $3\times 10^{-3}$, 
the evolution is essentially the same as those of the models with 
low $x_{\rm H_2, 0}$ (see \S 3.2 of Paper II).
Since $n_{c,0}$ is already higher than $n_{\rm H_2,cr}$
 and $n_{\rm HD,cr}$, the temperature increases as the collapse proceeds.
Therefore, even if HD cooling is more effective
than H$_2$ cooling at the initial state, 
H$_2$ cooling becomes dominant as the collapse proceeds.
In other words, the evolution is essentially
 the same as that without HD cooling.

The contraction time does not become longer than the fragmentation time
until the H$_2$ lines become optically thick
($n_c \sim 10^{12}$ cm$^{-3}$) due to
 the rapid increase in the H$_2$ abundance via
 the effective three-body reactions.
Therefore, the fragmentation is not expected to take place until the
cloud becomes opaque to the H$_2$ lines.

The evolution of the other models with different parameters 
($n_{\rm c,0} \gtrsim $ $10^5$ cm$^{-3}$,
 200 K $\lesssim T \lesssim $ 400 K, $3\lesssim f \lesssim 6$, and
 $10^{-4} \lesssim x_{\rm H_2, 0}\lesssim 10^{-2}$)
 is qualitatively similar to that of this model.

It is noted that for the high-density filaments with {\it low} $f$
 (e.g., $f\lesssim 2$ for $T_0 = 300$ K), 
 $t_{\rm dyn}$ becomes comparable to or longer than $t_{\rm frag}$ before
 $n_c$ reaches 10$^8$ cm$^{-3}$.
This is because when the parameter $f$ is small
 (or the initial line mass is small), 
 the equilibrium line mass of the filament, which is  
 proportional to the temperature, approaches the total line mass of the cloud
 before the three-body reactions become significant.
The contraction of such a cloud is decelerated significantly around
 $n_c \sim 10^8$ cm$^{-3}$, and thus the cloud is likely to
fragment around this density.

\section{Dependence of the Fragment Mass on the Initial Model Parameters}
\label{sec:dependence}

In this section, we estimate the fragment masses with our numerical results.
As shown above, there is a threshold H$_2$ abundance,
beyond which the HD cooling can play a key role in the thermal evolution of 
primordial gas clouds, and also there is a threshold initial density,
below which the filament would fragment at the critical density of 
H$_2$ or HD. Hence, as anticipated from the above numerical results,
the fragment mass is also divided into three cases, 
depending upon the initial H$_2$ abundance and density.

Comparing one-dimensional results 
with the two-dimensional results shown in Paper II,
we found that the fragmentation takes place when
$t_{\rm dyn}$ reaches 2$-$3 times the inverse of
the growth rate of the fastest-growing perturbation.
We thus assume that a filament fragments into dense cores when
the dynamical time reaches the fragmentation time,
$t_{\rm dyn} = t_{\rm frag}$.
The fragment masses are computed as $M_{\rm frag}\equiv m_{\rm eff}
\lambda$, where $m_{\rm eff}= \int 2\pi r \rho dr$ is a line mass
which is obtained by integrating the density
from the center to the radius at which the density 
takes 0.1$n_c$ and $\lambda$ is a longitudinal wavelength
of a fastest-growing linear perturbation
(eq.[38] of Nakamura et al. 1993).
This mass is about ten times larger than the Jeans mass.

\begin{figure}
\epsscale{1.0}
\plottwo{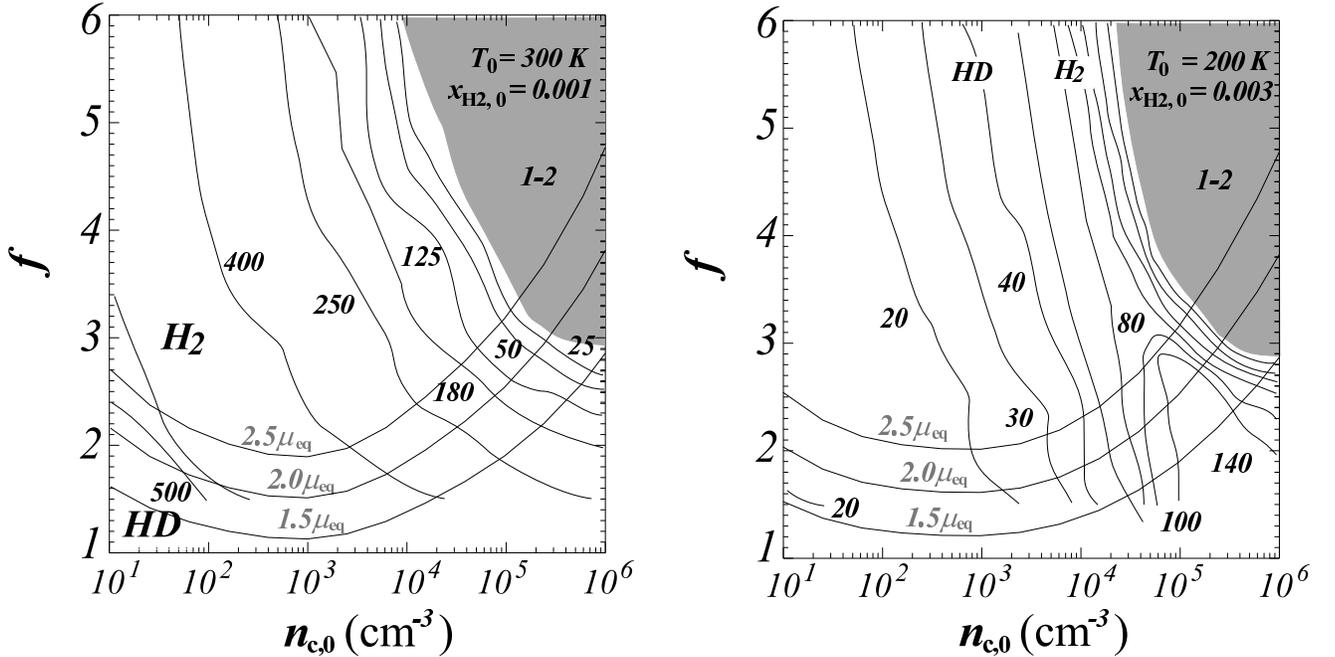}{f5b.ps}
\caption{The distributions of fragment mass for the models
 with (a) $T_0 = 300$ K and $x_{\rm H_2, 0} = 1\times 10^{-3}$ and
 (b) $T_0 = 200$ K and $x_{\rm H_2, 0} = 3\times 10^{-3}$.
($T_0$ is close to the equilibrium temperature for each value of 
$x_{\rm H_2, 0}$.)
The abscissa and ordinate denote the initial central density 
and the parameter $f$, respectively.
The solid lines denote the contours of the fragment mass which are
labeled with adjacent numbers in units of M$_\odot$.
A thick dashed line shows the lines at which 
$\Lambda _{\rm HD} = \Lambda _{\rm H_2}$
at the epoch of fragmentation.
In the left regions of the dashed lines, HD cooling is more 
efficient than H$_2$ cooling.
The dot-dashed lines show the line masses of the filaments
 formed by fragmentation of self-gravitating gas disks
 (see the text for more details), where $\mu _{\rm eq}$ 
 is a line mass of an equilibrium filament.
The number attached by each line denotes the line mass
 in units of $\mu _{\rm eq}$.
\label{fig:5}}
\end{figure}

Figures \ref{fig:5}a and \ref{fig:5}b show the fragment mass
for the models with $x_{\rm H_2, 0} = 1\times 10^{-3}$ and 
 $3\times 10^{-3}$, with assuming equilibrium initial temperature.
The abscissa and ordinate denote the initial central density 
and the parameter $f$, respectively.
The solid lines denote the contours of the fragment mass with
adjacent numbers in units of $M_\odot$.
The thick dashed lines show the boundary at which the HD cooling rate
is equal to H$_2$ cooling rate at the epoch of fragmentation.
In the regions to the left of the dashed lines, HD cooling is more 
effective than H$_2$ cooling.
Filamentary clouds are expected to form
 by the fragmentation of self-gravitating disks.
According to numerical simulations, 
when the filaments form by the growth of the
 most unstable modes, their line masses tend to be twice the line mass
for hydrostatic equilibrium, $\mu _{\rm eq} \equiv 2c_s^2/G$, where 
 $c_s$ is a sound speed.
For comparison, the line masses of the filaments with $1.5 \mu _{\rm
eq}$, $2 \mu _{\rm eq}$, and $2.5 \mu _{\rm eq}$ are shown by
dot-dashed lines.  The line mass of the equilibrium filament depends
only on the equilibrium temperature
 that is determined by the balance between the cooling time
 and the fragmentation time of the self-gravitating disk.  Here, the
fragmentation time is defined as the inverse of
 the most unstable linear perturbation (see Larson 1985).
The total line masses of the filaments are proportional to $fT_0$.
Therefore, the ordinate in Figure 5 corresponds to the total line 
masses of the filaments.

For the models with $x_{\rm H_2, 0} < x_{\rm H_2, cr} \approx 
3\times 10^{-3}$ (Fig. \ref{fig:5}a), 
the distribution of the fragment mass is quite similar to the case
without HD (Paper II), because HD cooling does not play an
important role in the thermal evolution of the filaments.
For higher initial density and/or larger $f$, the fragment mass is lower. 
The maximum and minimum masses are estimated as $\sim 10^3$M$_\odot$ 
and $1-2M_\odot$, respectively.  
The former corresponds to the Jeans mass at the stage at which the 
density reaches a critical density of H$_2$. 
The latter corresponds to the Jeans mass at the
stage at which the cloud becomes opaque to the H$_2$ lines. 
[The maximum mass is a few times smaller than that of Paper II
 because of the higher initial H$_2$ abundance.  
For the models with $x_{\rm H_2, 0} = 1\times 10^{-4}$, 
 the mass distribution as well as the maximum mass is in good agreement 
 with that of Paper II. See Figure 1a of Nakamura \& Umemura (2002).]

There is a steep boundary at $n_{\rm c,0}\sim 10^4 - 10^5$ cm$^{-3}$ 
in the distribution of the fragment mass for $f\gtrsim 3$. 
For the models with $n_{\rm c,0} \gtrsim 10^5$ cm$^{-3}$, 
the fragment masses take their minima at $1\sim 2$ M$_\odot$,
 whereas, for the models with $n_{\rm c,0} \lesssim 10^5$ cm$^{-3}$,
 they are greater than $\sim 10^2$ M$_\odot$.
This sensitivity in the fragment mass comes from 
the rapid increase in the H$_2$ abundance due to
 the three-body reactions.
For the models with low densities ($n_{\rm c,0}\lesssim 10^5$ cm$^{-3}$), 
 the contraction becomes quasistatic when the density reaches the
 critical density of H$_2$ and then linear density fluctuations can 
 grow nonlinearly before the three-body H$_2$ formation becomes
 dominant ($n_{\rm c,0}\gtrsim 10^{8-9}$ cm$^{-3}$).
In contrast, for the models with high densities
 ($n_{\rm c,0}\gtrsim 10^{5-6}$ cm$^{-3}$ and $f\gtrsim 3$),
the contraction time does not exceed the fragmentation time
 until the H$_2$ lines become optically thick
at $n\sim 10^{12}-10^{13}$ cm$^{-3}$.

On the other hand, when the initial H$_2$ abundance is higher than 
$3 \times 10^{-3}$ (Fig. \ref{fig:5}b), 
the HD cooling is more effective than the H$_2$
cooling for low-density filaments.
Thus, the maximum mass of the low-density region is reduced to
a few tens $M_\odot$ 
(The fragment mass depends weakly upon the initial H$_2$ abundance.
For example, for the models with $x_{\rm H_2, 0}= 1\times 10^{-2}$,
 the fragment mass is around 10 $M_\odot$ 
 in the low-density region.).
The reduced maximum mass is then related to the Jeans mass 
at the stage at which the density reaches a critical density of HD.
The minimum mass does not change because HD is not
a dominant coolant in the evolution of dense filaments.
Therefore, similarly to the models with low H$_2$ abundance,
the dependence of the fragment mass on the initial density
 exhibits a steep boundary
 around $n_{\rm c,0} = 10^4 - 10^5$ cm$^{-3}$ for $f\gtrsim 3$.

As mentioned in \S \ref{subsec:HighD}, for the high density filaments
 with low $f$, the contraction time temporally becomes longer than the
 fragmentation time just before the three-body reactions become important.
Thus, fragmentation is expected to take place before the three-body
 H$_2$ formation becomes efficient.
 As a result, the fragment masses are relatively large
 (the lower-right regions in Figure \ref{fig:5}).


In summary, the distributions of the fragment mass
has a steep boundary around $n_{\rm c,0}\gtrsim 10^{4-5}$ cm$^{-3}$
for $f\gtrsim 3$, irrespectively of the initial H$_2$ abundance.
This implies that the IMFs in very metal-deficient gas are
 likely to be bimodal if both low-density and high-density filaments
form stars.
The low-mass peak is around a few $M_\odot$, which is not sensitive to
the abundance of H$_2$ formed in a parent cloud.
The high-mass peak is $\approx 10^2 M_\odot$
 if $x_{\rm H_2,0} \lesssim 3\times 10^{-3}$, while it is
$\approx 10 M_\odot$ if $x_{\rm H_2,0} \gtrsim 3\times 10^{-3}$.

\section{Discussion}


First, we briefly discuss the role of HD for the
formation of the very first stars (Pop III stars).
In the bottom-up scenarios like cold dark matter models,
 when the first pregalactic objects with masses of $10^5 - 10^8$
 M$_\odot$ are virialized at redshifts of $z\sim 10 - 10^2$,
 the H$_2$ abundance reaches at most $10^{-4} - 10^{-3}$
 which is lower than the threshold H$_2$ abundance.
Therefore, the cloud evolution is basically determined by H$_2$ cooling
rather than HD cooling (Nakamura \& Umemura 1999b, 2000).
If the primordial D abundance is a few times as high as the value we assumed
 ($x_{\rm D}=4\times 10^{-5}$), HD cooling may play a role
 in the thermal evolution of the pregalactic clouds.
However, recent observations of quasar absorption spectra find an
observed primordial D abundance as low as $3-4\times 10^{-5}$
(Tytler et al. 1996), implying that HD cooling is not important for the
first star formation.  

Next, we give some implications for star formation 
in metal-deficient primordial galaxies.
Recently, Susa \& Umemura (2000) investigated the pancake collapse
 of pregalactic clouds under UV background radiation.
They found that once the pancake disk is shielded against
 external UV radiation in the course of contraction,
 H$_2$ molecules form efficiently via the H$^-$ reaction
 with abundant free electrons produced by
 UV background, and the resultant abundance reaches 
$x_{\rm H_2} \approx 3\times 10^{-3}$
 (see also Shapiro \& Kang 1986).
The pancake disks probably fragment into filaments in which 
stars can subsequently form.
In this case, HD cooling is expected to become efficient
 in low-density filaments, and then
 the high mass peak of the IMF would go down to
 $\sim$ 10 $M_\odot$.

It is found that at redshifts of $z\sim 2$,
 the UV background radiation decreases with time 
 (Irwin, McMahon, \& Hazard 1991; Maloney 1993).
The time-decreasing UV background radiation is likely to 
 influence star formation in galaxies,
 especially low surface brightness galaxies 
 (e.g., Ellis 1997).  
Corbelli et al. (1998) studied the effects of
 the declined UV background radiation
 on the thermal evolution of the protogalaxies
 with low surface densities.  
They found that there is a critical redshift
 of $z \sim 1-2$, above which  
 the gas disks with surface densities 
$10^{20}$ cm$^{-2} \lesssim N_{\rm HI} \lesssim 10^{21}$ cm$^{-2}$
 are gravitationally stable at $T\sim 10^4$ K.
Below this redshift, the declined UV radiation is shielded by the gas
 disks where the H$_2$ abundance reaches $10^{-2}$ owing to high
 ionization degree by the UV radiation.
Also, in such galaxies, the high mass peak of the IMF 
 would decrease to $\sim 10M_\odot$ owing to the HD cooling.

The high mass end of the IMF can influence 
abundance patterns because metal production by
extremely metal-deficient stars 
is very different between $10^2 M_\odot$ and $10 M_\odot$ 
(Abia et al. 2001; Umeda \& Nomoto 2001; Heger \& Woosley 2001; Schaerer 2001).
For instance, the abundance pattern in the metal-poor ($\sim 0.05 Z_\odot$) 
starburst galaxy M82 cannot be accounted for
unless stars with $\gtrsim 25 M_\odot$
contribute  significantly to the metal enrichment of the galaxy
(Tsuru et al. 1997; Nakamura et al. 2001; Umeda \& Nomoto 2001). 
This mass scale seems to be consistent with
 the high mass peak of the present IMF regulated by the HD cooling.
It is also consistent
 with the estimate by Hernandez \& Ferrara (2001).

As the star formation progresses,
 the interstellar metallicity will monotonously
 increase with time.  When the metallicity reaches 
$10^{-3} - 10^{-2}Z_\odot$, the metal cooling becomes important 
and the thermal properties of the gas are changed.
Thereafter, the process of star formation would 
become similar to the present-day case.
In other words, the IMF would settle into Salpeter-like IMF.

\acknowledgments

We are grateful to A. Ferrara, T. Nakamoto, R. Nishi, K. Omukai, H. Susa,
 and H. Uehara for stimulating discussion.
Numerical computations 
 were carried out on VPP300/16R 
 at the Astronomical Data Analysis Center
 of the National Astronomical Observatory, Japan and 
 on workstations at the Center for Computational Physics, 
 University of Tsukuba.
This work was financially supported in part by the Grant-in-Aid 
 for Scientific Research on Priority Areas 
 of the Ministry of Education, Science, Sports and Culture
 10147205 and 11134203 (FN).

\begin{deluxetable}{llll}
\tablecolumns{4}
\tablecaption{REACTION RATE COEFFICIENTS}
\tablehead{
&\colhead{Reactions} & \colhead{Rate Coefficients
 (cm$^3$s$^{-1}$)} 
& \colhead{Reference}}
\startdata
  (D1) & ${\rm D^+}+ e \rightarrow {\rm D}+h\nu$
& $k _{D1} = 3.6 \times 10^{-12} (T/300)^{-0.75}$ & SLD98 \\
  (D2) & ${\rm D}+ {\rm H^+} \rightarrow {\rm D^+}+{\rm H}$
& $k _{D2} = 3.7 \times10^{-10} T^{0.28}
 \exp \left(-43/T\right) $  & GP98 \\
  (D3) & ${\rm D^+}+ {\rm H} \rightarrow {\rm D}+{\rm H^+}$
& $k _{D3} = 3.7 \times 10^{-10}T^{0.28}$ & GP98 \\
  (D4) & ${\rm D}+ {\rm H} \rightarrow {\rm HD}+h\nu$
& $k _{D4} = 1.0 \times10^{-25}$ & GP98 \\
  (D5) & ${\rm D}+ {\rm H_2} \rightarrow {\rm H}+{\rm HD}$
& $k _{D5} = 9.0 \times 10^{-11} \exp \left(-3876/T\right)$
& GP98 \\
  (D6) & ${\rm HD^+}+ {\rm H} \rightarrow {\rm H^+}+{\rm HD}$
& $k _{D6} = 6.4 \times 10^{-10}$ & GP98 \\
  (D7) & ${\rm D^+}+ {\rm H_2} \rightarrow {\rm H^+}+{\rm HD}$
& $k _{D7} = 2.1 \times 10^{-9}$ & GP98 \\
  (D8) & ${\rm HD}+ {\rm H} \rightarrow {\rm H_2}+{\rm D}$
& $k _{D8} = 3.2 \times 10^{-11} \exp \left(-3624/T\right)$
& GP98 \\
  (D9) & ${\rm HD}+ {\rm H^+} \rightarrow {\rm H_2}+{\rm D^+}$
& $k _{D9} = 1.0 \times 10^{-9} \exp \left(-464/T\right)$ & GP98 \\
  (D10) & ${\rm D}+ {\rm H^+} \rightarrow {\rm HD^+}+h\nu$
& $k _{D10} = {\rm dex} [-19.38 -1.523\log T+$ & \\
& & \hspace{1cm} $1.118(\log T)^2 -0.1269(\log T)^3]$ & GP98 \\
  (D11) & ${\rm D^+}+ {\rm H} \rightarrow {\rm HD^+}+h\nu$
& $k _{D11} = {\rm dex} [-19.38 -1.523\log T+ $ & \\
& & \hspace{1cm} $1.118(\log T)^2 -0.1269(\log T)^3]$ & GP98 \\
  (D12) & ${\rm HD^+}+ {\rm e} \rightarrow {\rm H}+{\rm D}$
& $k_{D12}=7.2 \times 10^{-8}T^{-1/2}$ & GP98 \\
  (D13) & ${\rm D}+ e \rightarrow {\rm D^-}+h\nu$
& $k_{D13}=3.0 \times 10^{-16}(T/300)^{0.95}\exp(-T/9320)$ & SLD98 \\
  (D14) & ${\rm D^+}+ {\rm D^-} \rightarrow 2{\rm D}$
& $k_{D14}=5.7 \times 10^{-8}(T/300)^{-0.5}$ & SLD98 \\
  (D15) & ${\rm H^+}+ {\rm D^-} \rightarrow {\rm D}+{\rm H}$
& $k_{D15}=4.6 \times 10^{-8}(T/300)^{-0.5}$ & SLD98 \\
  (D16) & ${\rm H^-}+ {\rm D} \rightarrow {\rm H}+{\rm D^-}$
& $k_{D16}=6.4 \times 10^{-9}(T/300)^{0.41}$ & SLD98 \\
  (D17) & ${\rm D^-}+ {\rm H} \rightarrow {\rm D}+{\rm H^-}$
& $k_{D17}=6.4 \times 10^{-9}(T/300)^{0.41}$ & SLD98 \\
  (D18) & ${\rm D^-}+ {\rm H} \rightarrow {\rm HD}+e$
& $k_{D18}=1.5 \times 10^{-9}(T/300)^{-0.1}$ & SLD98
\enddata
\tablecomments{GP98: Galli \& Palla (1998); 
SLD98: Stancil, Lepp, \& Dalgarno (1998) 
}
\label{tab:hd rate}
\end{deluxetable}

\clearpage

\end{document}